\documentclass[twocolumn,pra]{revtex4}
\usepackage{graphicx}
\newcommand{\ket}[1]{|{#1}\rangle}			
\newcommand{\bra}[1]{\langle{#1}|}

\begin{document}
\title{Unconditional security of the Bennett 1992 quantum key-distribution scheme with strong reference pulse}

\author{Kiyoshi Tamaki$^{1,2}$}
\email{tamaki@will.brl.ntt.co.jp}
\author{Norbert L\"utkenhaus$^{3,4}$}
\author{Masato Koashi$^{2,5}$}
\author{Jamie Batuwantudawe$^{3}$}
\affiliation{$^{1}$NTT Basic Research Laboratories, NTT Corporation,\\
3-1,Morinosato Wakamiya Atsugi-Shi, Kanagawa, 243-0198, Japan\\ 
$^{2}$CREST, JST Agency, 4-1-8 Honcho, Kawaguchi, Saitama, 332-0012, Japan\\
$^{3}$Institute for Quantum Computing, University of Waterloo, Waterloo, ON, N2L 3G1, Canada\\
$^{4}$Quantum Information Theory Group, Institut f\"{u}r Theoretische Physik,
Universit\"{a}t Erlangen-N\"{u}rnberg, Staudtstr. 7/B3, 91058 Erlangen, Germany\\
$^{5}$Division of Materials Physics, Department of Materials Engineering Science\\
Graduate school of Engineering Science, Osaka University, Toyonaka, Osaka 560-8531, Japan
}

\begin{abstract}
We prove the unconditional security of the original Bennett 1992 protocol with strong reference pulse. We show that we may place a projection onto suitably defined qubit spaces before the receiver, which makes the analysis as simple as qubit-based protocols. Unlike the single-photon-based qubits, the qubits identified in this scheme are almost surely detected by the receiver even after a lossy channel. This leads to the key generation rate that is proportional to the channel transmission rate for proper choices of experimental parameters.

\end{abstract}

%\pacs{03.67.Dd, 03.67.-a}

%\date{\today}

\maketitle
%%%%%%%%%%%%%Text%%%%%%%%%%

%%%%%%%%%%%%%%%%%%%%%%%%%%%%%%%%%%%%%%%%%%%%%%%%figure1
\begin{figure}[tbp]
\begin{center}
 \includegraphics[scale=0.3]{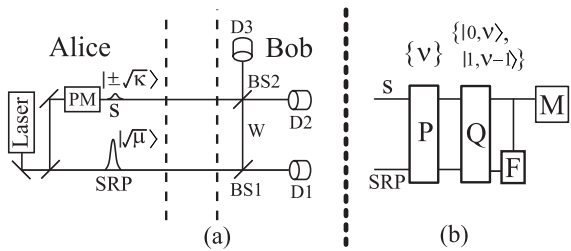}
\end{center}
 \caption{(a) An experimental setup for the B92 protocol with strong reference pulse. PM: phase modulator. (b) An equivalent setup for Bob.
\label{setup}}
\end{figure}
%%%%%%%%%%%%%%%%%%%%%%%%%%%%%%%%%%%%%%%%%%%%%%%%

\section{Introduction}
Quantum key distribution (QKD) is an art to distribute a secret key between parties (Alice and Bob) with arbitrary small leakage of its information to an unauthorized party (Eve). So far, several QKD protocols have been proposed and experimentally demonstrated \cite{DLM06}. In real life QKD, loss and noises in the quantum channel limit the achievable distance. In order to cover longer distances, a decoy state idea for BB84 \cite{BB84} with weak coherent pulses was proposed, which achieves a key generation rate of order $O(\eta)$ \cite{W03} in the single photon transmission rate $\eta$, whereas it is in the order of $O(\eta^2)$ without decoy states \cite{ILM01}. 

Another proposal to achieve longer distances was made in 1992 by Bennett
\cite{B92} (B92). The protocol uses two weak coherent states together with a strong reference pulse (SRP), and is expected to be robust
 against channel losses. In contrast to the decoy schemes where the robustness is achieved 
by using an increased number of states, the B92 protocol uses just 
two states, and hence its robustness 
should come from an entirely different mechanism.
Unconditional security was proved  \cite{TL04}
for a variant of the B92 protocol
using two polarization states of a single photon,
which we call S-B92,
but it does not share the robustness expected 
for the original B92.
 The security proof of the original B92 protocol is 
thus important from the practical viewpoint
as well as the fundamental one.

In this paper, we prove the unconditional security of the original
B92 protocol, and reveal the mechanism with which 
its robustness against loss arises. The crucial finding is that 
 the B92 protocol can be regarded as many S-B92 protocols executed in parallel,
which has two benefits. Firstly, it allows us to use the analysis 
of S-B92 in proving the security of the B92 protocol. Secondly,
a striking difference from S-B92 tells us why the B92 protocol is robust against loss and performs similarly to an ideal
single-photon implementation of the BB84 protocol.

In two-state protocols, it is known that by exploiting an unambiguous state
discrimination (USD) measurement, Eve can obtain a benefit from qubit
loss events \cite{DLM06}. Thus, a secure key is impossible over longer
distances in S-B92 \cite{TL04}, where qubit loss events are directly
connected to physical channel loss. 
On the other hand, the qubits in the B92 protocol turn out to be defined including a vacuum component in the signal mode,
and they are almost always received 
by Bob even after traveling over longer distances.
In other words, the qubit loss events are
negligible since it is {\it not} directly connected
to the physical channel loss.
% This is the main intuition we obtained in this paper.

%This paper is organized as follows. First, we show how the B92 protocol works. Secondly, we briefly review S-B92 and its security proof. Then, we convert the B92 protocol to EDP-B92, where we will see that the B92 protocol can be regarded as running many S-B92 in parallel. Next, we mention how to prove the security of EDP-B92, and finally we show some security examples.

This paper is organized as follows. In Sec. II, we explain how the B92 protocol works, and in Sec. III, we prove its security. Our security proof is based on the conversion to an entanglement distillation protocol, and our proof uses technique from the security proof of S-B92. Thus, in Sec. III A, we briefly review the security proof of S-B92 \cite{TL04}, then in Sec. III B, III C, and III D, we intuitively present the security proof of the B92 protocol, putting some emphasis on the difference between S-B92 and the B92 protocol. We leave the technical details of the proof to the appendices. Finally, we show some examples of the resulting key generation rate in Sec. IV, and we summarize this paper in Sec. V.

\section{B92 protocol with strong reference pulse}

The essence of the experimental setup of the B92 protocol can be expressed by a
Mach-Zehnder interferometer in Fig.~\ref{setup}(a). Alice generates a
strong coherent pulse and splits it into a weak signal pulse
$\ket{\sqrt{\kappa}}_{\rm S}$ and an SRP $\ket{\sqrt{\mu}}_{\rm SRP}$
($\kappa\ll\mu$). After Alice applies a phase modulation to prepare
$\ket{(-1)^i\sqrt{\kappa}}_{\rm S}$ according to her random bit $i=0,1$,
she sends out these two systems. Bob uses a beam
splitter BS1 with reflectivity $R\equiv\kappa/\mu$, which splits the
SRP into weak and strong pulses. The weak pulse (W) and the signal pulse
interfere at BS2, whose action 
is represented by $\hat{b}_{j}=\frac{1}{\sqrt{2}}\left(\hat{a}_{\rm
W}+(-1)^{j}\hat{a}_{\rm S}\right) (j=2,3)$, where $\hat{b}_{j}$ and
$\hat{a}_{\rm W/S}$ are annihilation operators for the spatial mode to
the detector ${\rm D}_{j}$ and the W/S mode, respectively.

For Bob's part, we assume two types of detectors. One type can tell if the photon number $\nu$ is in an interval 
\begin{eqnarray}
\lambda^{({\rm D}_1)}\equiv [\nu_{i},\nu_{f}-1],
\label{D1}
\end{eqnarray} 
and the other type can discriminate among vacuum, single-photon, and multi-photon events. We note that one of us \cite{K04} proved the security of a
modified B92 with threshold detectors, but there Bob must lock his own 
local oscillator to the SRP mode via a feed-forward control, which  
may not be easy to implement. 
In Fig.~\ref{setup}(a), Bob infers Alice's bit value to be $0$ $(1)$ when ${\rm D}_2$ $({\rm D}_3)$
 records a single photon and ${\rm D}_3$ $({\rm D}_2)$ records no photon.
Let $\Lambda_{\rm fil,all}$ be the rate of these conclusive 
events. Here and henceforth, ``rate'' is always normalized by the total number of signals. Among the conclusive events, only the cases where the outcome of 
${\rm D}_1$ is 
in $\lambda^{({\rm D}_1)}$ will be kept
and used to generate the final
key after classical error correction (EC) and privacy amplification
(PA) \cite{nielsen}.
Let $\Lambda_{{\rm fil},\lambda'}$ be the rate of these events, 
where 
\begin{eqnarray}
\lambda'\equiv[\nu_i+1,\nu_f]
\label{lambdaprime}
\end{eqnarray}
corresponds to
 the range of the total number of photons recorded by the three detectors.
Using random test bits, Alice and Bob monitor the error rates 
$\Lambda_{\rm bit,all}$ and $\Lambda_{{\rm bit},\lambda'}$ 
of the cases where Alice's bit and Bob's bit are different.
Bob also monitors the rate $\Lambda_{{\rm vac},\lambda^{({\rm D}_1)}}$
of events where both ${\rm D}_2$ and ${\rm D}_3$ record no photon
and ${\rm D}_1$ reports an outcome in $\lambda^{({\rm D}_1)}$. Note that in the normal operation with
 $\eta\ll 1$ and $\lambda^{({\rm D}_1)}$ being wide enough, 
we should have $\Lambda_{\rm fil,all}\cong \Lambda_{{\rm
fil},\lambda'}$, $\Lambda_{\rm bit,all}\cong \Lambda_{{\rm
bit},\lambda'}$, and 
$\Lambda_{{\rm vac},\lambda^{({\rm D}_1)}}\cong \Lambda_{{\rm vac,all}}
\cong 1$, but these three rates must still be monitored to watch out for Eve's possible attacks as we will see later.

\section{Security proof}
In this section, we prove the security of the B92 protocol. Throughout this paper, we assume that all imperfections are controlled
by Eve. Note that ${\rm D}_1$ with quantum efficiency $\eta_1$ and
${\rm D}_2$ (${\rm D}_3$) with $\eta'$ are equivalent to the
setup with unit-efficiency detectors when 
absorbers with the transmission rate of $\eta'$ and
$\eta''=R\eta'+(1-R)\eta_1$ are respectively put in S and
SRP modes, and the reflectivity of BS1 is changed to
$R\eta'/(R\eta'+(1-R)\eta_1)$. As for dark counts, we assume Eve can induce dark counts as she pleases. Thus, in what follows we assume that all detectors have unit efficiency and no dark counts.

Since the B92 protocol and S-B92 share many similarities, in the security proof of the B92 protocol we employ the idea from security proof of S-B92. In the next subsection, we briefly review S-B92 security proof. It is based on the conversion of the protocol to an entanglement distillation protocol. After the review, we present the security proof of the B92 protocol.

\subsection{Review of B92 with single-photon implementation}
\label{S-B92}

In S-B92, Alice sends out $\ket{\varphi_{i}}\equiv\beta\ket{0_x}+(-1)^i\alpha\ket{1_x}$ ($\alpha^2+\beta^2=1$ and $\beta>\alpha>0$) depending on randomly chosen bit value $i=0,1$, where $\ket{0_x}$ and $\ket{1_x}$ represent a basis (X-basis) of the qubit states (the single-photon polarization states). For later convenience, we define the Z-basis as $\ket{i_z}\equiv(\ket{0_x}+(-1)^i\ket{1_x})/\sqrt{2}$, and $\hat{P}(\ket{\phi})\equiv\ket{\phi}\bra{\phi}$. When Bob receives a qubit state, he broadcasts this fact and continues to perform a measurement described by positive-operator-valued measure (POVM) \cite{nielsen} 
\begin{eqnarray}
\hat F_0&=&(G/2)\hat P(\ket{\overline{\varphi}_1})\nonumber\\
\hat F_1&=&(G/2)\hat P(\ket{\overline{\varphi}_0})\nonumber\\
\hat F_{\rm null}&=&1-\hat F_{0}-\hat F_{1}\,,
\end{eqnarray}
where $G=(\gamma/\beta)^2$, $0<\gamma\le1$, and $\ket{\overline{\varphi}_i}\equiv\alpha\ket{0_x}-(-1)^i\beta\ket{1_x}$ is the state orthogonal to $\ket{\varphi_i}$. We call measurement outcomes with $F_0$ or $F_1$ conclusive. Bob tells over a public channel if he has obtained the conclusive event or not. 

In order to prove the security of S-B92, in \cite{TL04} S-B92 was converted to an entanglement distillation protocol (EDP). A key point in the conversion
is that Bob's measurement
can be decomposed by a filtering operation \cite{G96}, whose successful
operation is represented by a Kraus operator \cite{nielsen} 
\begin{eqnarray}
\hat A_{\rm s}=\sqrt{G}[\alpha\hat P(\ket{0_x})+\beta\hat P(\ket{1_x})]
\end{eqnarray}
followed by
Z-basis measurement. This can be seen by noting that $\hat F_i=\hat P(A_{\rm
s}\ket{i_z})$, and the successful filtering operation corresponds to the
conclusive event in S-B92. 
On the other hand, we assume that Alice prepares two qubits in
a state $\ket{\Phi}_{\rm s}\equiv\frac{1}{\sqrt{2}}(\ket{0_z}_{\rm
A}\ket{\varphi_0}_{\rm B}+\ket{1_z}_{\rm A}\ket{\varphi_1}_{\rm B})$,
sends only qubit B to Bob, 
and then performs Z-basis measurement on qubit A. In the normal
operation without Eve, the successful filtering operation orthogonalizes
$\ket{\varphi_0}_{\rm B}$ and $\ket{\varphi_1}_{\rm B}$ so that they
share a maximally entangled state, 
and the bit value shared by Alice and Bob via Z-basis measurements is secure.

In the presence of loss, noise, or Eve's intervention, 
Alice and Bob do not share a maximally entangled state,
but if they succeed in estimating the bit and the phase error rate 
on the qubit pairs from which they create the key, then they can distill an almost perfect maximally entangled state by running an EDP. Here, the bit
(phase) error stands for an event where Alice and Bob have different 
measurement outcomes in Z-basis (X-basis). Moreover, if we use the argument by Shor and Preskill \cite{SP00}, the EDP protocol followed by Z-basis measurements can be made equivalent to S-B92 with EC and PA.
Since the bit error rate $\Lambda_{{\rm bit}}$ 
can be reliably estimated from test bits, 
we are only left with the estimation of the upper bound on the phase error
rate $\Lambda_{{\rm ph}}$ from the observed variables. Thanks to the filter, we can actually relate this bound with other observables. When the photon reaches Bob at rate $\Lambda_{\rm s}$
and gives a conclusive outcome at rate $\Lambda_{\rm fil}$,
it was shown in \cite{TL04} that
\begin{equation}
 \Lambda_{{\rm fil}}-2\Lambda_{{\rm bit}}\le
2G\,\alpha\,\beta\,g(C{\bf Z})\, ,
\label{sfinal}
\end{equation}
where $g((a,b,c,d)^T)\equiv \sqrt{ad}+\sqrt{bc}$, 
${\bf Z}=(\Lambda_{\rm s}, \Lambda_{1x}, \Lambda_{\rm fil},
\Lambda_{{\rm ph}})^T$, and the matrix $C$ is the inverse of
\begin{equation}
C^{-1}=\left(
\begin{array}{cccc}
1&1 &1 &1 \\
0&0 &1 &1 \\
G\alpha^2&G\beta^2 &G\alpha^2 &G\beta^2 \\
0&G\beta^2 &G\alpha^2 & 0
\end{array}
\right)\,.
\end{equation}
The quantity $\Lambda_{1x}$ is the rate at which Bob receives 
the photon and Alice's qubit A is in state $\ket{1_x}_{\rm A}$.
Although $\Lambda_{1x}$ is not measured in the protocol,
 we have a bound $\alpha^2-1+\Lambda_{\rm s}\le\Lambda_{1x} \le \alpha^2$. 
Using this, the bound on $\Lambda_{{\rm ph}}$ can be obtained by solving 
Eq.~(\ref{sfinal}).

%, but care must be taken since the rate 
%$\Lambda_{1x}$ is not directly measured in the protocol. 
%A simple channel loss would result in $\Lambda_{\rm s}=\eta$ and 
%$\Lambda_{1x}=\eta\beta^2$, but since 
%Eve can freely determine which of the photons should be detected by Bob,
%we must assume that $\Lambda_{1x}$ can take
% any value in the range $[\beta^2-(1-\eta),\beta^2]$. 

The key rate resulting from the above formula is nonzero only for 
 short distances \cite{TL04}. 
For a high loss region with small rate $\Lambda_{\rm s}$, 
Eve may perform the USD measurement on the incoming two states.
Whenever it fails, she blocks the photon as if it were lost 
due to the channel loss. If the USD is successful,
she can send the correct state to Bob and hence 
she learns the bit value without inducing any error.
It follows that {\it any} two state protocol is fragile against qubit loss events. Note that the qubit loss events is not always connected directly to the physical channel losses, which is the case for the B92 protocol as we will see.

%One can see its reason as follows. Imagine that Eve implements an USD measurement on the two states, which succeeds with some probability. If she gets the successful measurement outcome, then she resends the corresponding state, otherwise she sends a vacuum state, which corresponds to qubit loss events. Note that this eavesdropping induces no bit errors and gives her complete information on the resending states so that Eve can break S-B92 assuming high enough channel losses. In other words, {\it any} two state protocol is fragile against qubit loss events. 

\subsection{Conversion of the B92 protocol into an entanglement distillation protocol}

In this subsection and the subsequent ones, we prove the security of the B92 protocol. Especially, this subsection is devoted to a conversion of the B92 protocol into an entanglement distillation protocol, which is the most crucial point in our proof.

Our strategy for the proof is to apply a similar argument to the original B92 with SRP, i.e., we consider the distillation of the maximally entangled state. For the distillation, we first have to convert Alice's and Bob's part in such a way that each of them has a qubit. Then, we consider the phase error estimation by modifying Eq.~(1). Alice's part can easily be converted by letting her first prepare $\ket{\Psi}\equiv\frac{1}{\sqrt{2}}\left(\ket{0_z}_{\rm
A}\ket{\sqrt{\kappa}}_{\rm S}+\ket{1_z}_{\rm A}\ket{-\sqrt{\kappa}}_{\rm
S}\right)\ket{\sqrt{\mu}}_{\rm SRP}$ and measure the qubit A on
the Z basis. 

On the other hand, the conversion of Bob's part is not straightforward since the definition of his qubit space is by no means trivial. For the conversion, we first introduce additional fictitious measurements which do not disturb Bob's conclusive data at all. Thus, the security of the final key does not change even if we assume that Bob has conducted these additional measurements. As we will see, these measurements consist of a total photon number measurement (P) and a qubit projection (Q). Furthermore, as we did in Sec.~\ref{S-B92} for the S-B92 protocol, we decompose Bob's original measurement into a filtering operation (F) followed by a projection measurement (M) (see Fig.~\ref{setup}(b)). Here, the qubit space is defined by combining the vacuum state and the single-photon state of the S-mode with the appropriate Fock states of the SRP mode depending on the outcome of measurement P. Our qubit space is thus defined depending on the total photon number. This total photon number determines the nonorthogonality in Bob's measurement.

%After the successful definition of the qubit space, we can estimate the phase error rate for each qubit space by using Eq.~(1) given experimental data for the qubit space. Note, however, that in the actual experiment, those experimental data for each qubit space is not available. Hence, we have to consider how to treat each qubit space jointly in such a way that the corresponding data is accessible in the actual experiment. Fortunately, thanks to the convexity and Azuma's inequality, we can actually estimate the phase error rate in the joint manner from the data available in the experiment.

Now, let us present the conversion of Bob's part in detail. First, define $\ket{\nu',\nu-\nu'}_{\rm B}\equiv \ket{\nu'}_{\rm S}\ket{\nu-\nu'}_{\rm SRP}$ as the state with $\nu'$ photons and $\nu-\nu'$ photons in mode S and SRP, respectively. Since Bob uses only linear optics and photon detectors, it does not disturb the statistics of Bob's measurement outcomes to assume that
Bob's measurement is preceded by the measurement of the total photon number $\nu$
in modes S and SRP. We denote this measurement as ``P'', whose POVM is
$\{\sum_{k+k'=\nu}\hat{P}(\ket{k,k'}_{\rm B})\}_{\nu=0,1,\cdots}$.
 Suppose that $\nu>0$ is obtained in this measurement, and recall that Bob's conclusive events occur only when 
D$_2$ and D$_3$ receive one photon in total. Then, it follows that without any disturbance of the statistics of the {\it conclusive} events, P can be followed by projection measurement ``Q'' with POVM $\{\sum_{i=0,1}\hat{P}(\ket{i,\nu-i}_{\rm B}), \sum_{\nu'=2,3,\cdots,\nu}\hat{P}(\ket{\nu',\nu-\nu'}_{\rm B})\}$. Here, Q decides if the state is in the qubit subspace ${\cal H}^{(\nu)}$ spanned by $\ket{0_x^{(\nu)}}_{\rm B}\equiv \ket{0,\nu}_{\rm B}$ and $\ket{1_x^{(\nu)}}_{\rm B}\equiv \ket{1,\nu-1}_{\rm B}$.

Under the condition that the total number of photons was $\nu$, let us calculate the POVM element $\hat{F}_i^{(\nu)}$ for obtaining conclusive bit $i$. At BS1, conversion $\ket{0,\nu}_{\rm B}\rightarrow\ket{0}_{\rm S}\ket{1}_{\rm W}\ket{\nu-1}_{\rm D_{1}}$ occurs with probability $(1-R)^{\nu-1}\nu R$, and $\ket{1,\nu-1}_{\rm B}\rightarrow\ket{1}_{\rm S}\ket{0}_{\rm W}\ket{\nu-1}_{\rm D_{1}}$ with $(1-R)^{\nu-1}$. Since outcome $i$ corresponds to the projection to state $(\ket{0}_{\rm S}\ket{1}_{\rm W}+(-1)^i\ket{1}_{\rm S}\ket{0}_{\rm W})/\sqrt{2}$, we have 
\begin{eqnarray}
\hat{F}_i^{(\nu)}=(G_\nu/2)\hat{P}
(\alpha_\nu\ket{0_x^{(\nu)}}_{\rm B}+(-1)^i \beta_\nu\ket{1_x^{(\nu)}}_{\rm B})
\label{Bit value}
\end{eqnarray}
with $G_\nu\equiv (1-R)^{\nu-1}(1+\nu R)$,
$\alpha_\nu\equiv \sqrt{\nu R/(1+\nu R)}$,
and $\beta_\nu\equiv \sqrt{1/(1+\nu R)}$.
At this point, we notice that the measurement on each qubit space is the same as the measurement in S-B92 with $(G,\beta,\alpha)\to
(G_{\nu},\beta_{\nu},\alpha_{\nu})$,
and hence we can define the filter accordingly.
The successful filtering operation converts
$\beta_\nu\ket{0^{(\nu)}_{x}}\pm\alpha_\nu\ket{1^{(\nu)}_{x}}$ into
orthogonal states, and it is described by 
\begin{eqnarray}
\hat A_{\rm s}^{(\nu)}=\sqrt{G_{\nu}}[\alpha_{\nu}\hat P(\ket{0_x^{(\nu)}})+\beta_{\nu}\hat P(\ket{1_x^{(\nu)}})]\,.
\label{filter}
\end{eqnarray}
Physically, the conversion is realized by 
BS1 and $(\nu-1)$-photon detection
by D$_1$, resulting in $(\ket{0}_{\rm S}\ket{1}_{\rm W}\pm\ket{1}_{\rm S}\ket{0}_{\rm W})/\sqrt{2}$. Thus, we conclude that Bob's measurement can be decomposed into the
sequence of P, Q, and the filtering operation (F) followed by 
a projection measurement on a qubit. 

Note that in the B92 protocol we distill a key from the successfully filtered events (or the conclusive events) with total photon number $\nu\in\lambda'$. Thus, by following the argument by Shor and Preskill, we conclude that once we have the estimation of an upper-bound on the phase error rate with total photon number $\nu\in\lambda'$, i.e., $\Lambda_{{\rm ph}, \lambda'}/\Lambda_{{\rm fil}, \lambda'}$, we complete the security proof.

%This ends the construction of an EDP protocol corresponding to the B92 protocol, i.e., EDP-B92.%, and if Alice and Bob can distill a MES from the system $A$ and successfully filtered Bob's qubit state, then they can share a secure key.

%It is instructive to see how $\ket{\pm\sqrt{\kappa}}_{\rm S}\ket{\sqrt{\mu}}_{\rm SRP}$ evolves under Bob's operations assuming that Eve is absent. The $\nu$-photon detection by P and the successful Q project the states to $\frac{1}{\sqrt{1+\nu R}}[\ket{0,\nu}\pm\sqrt{\nu R}\ket{1,\nu-1}]$. Then, Bob performs the filtering whose successful operation converts these two states into orthogonal states as $\frac{1}{\sqrt{2}}[\ket{0,1}\pm\ket{1,0}]$ so that Alice and Bob can share MES in the normal operation with some probabilities. At this point, we notice a similarity and differences of the B92 protocol from S-B92. 

\subsection{Comparision to the single-photon implementation}

S-B92 and our B92 are similar in the sense that Bob applies a filter
that orthogonalizes two nonorthogonal qubit states. However, the two
schemes differ in the  structure of the qubit spaces and the effect of optical loss (represented by channel transmission $\eta$) on them. 
In the S-B92 protocol, the loss of the photon means that Bob fails to receive a qubit. On the other hand, when the photon reaches Bob, the nonorthogonality of the two received states is the same as that of the two states released from Alice. Hence the initial states of Alice and the measurement by Bob are described by the same parameter $\alpha$. The loss of the qubit is fatal to the key rate, since Eve is able to let Bob receive a qubit preferentially in the cases where she has succeeded in discriminating the states in the USD measurement. 
In the B92 protocol, by contrast, Bob almost always receives a qubit ${\cal H}^{(\nu)}$: He only fails when $\nu$ is outside of $\lambda'$ or when Q fails the qubit projection, both of which are negligible if $\lambda'$ is chosen to be wide enough and $\eta\kappa\ll 1$. This prevents Eve from forcing Alice and Bob to accept the events favorable to her.

The effect of the optical loss in the B92 protocol shows up in an entirely different place. To see this, let us rewrite Alice's initial coherent states as
$\ket{\pm \sqrt{\kappa}}_{\rm S}=\tilde\beta \ket{+}_{\rm S}\pm 
\tilde{\alpha}\ket{-}_{\rm S}$ with 
\begin{equation}
\tilde{\alpha}^2=(1-e^{-2\kappa})/2,
\label{alpha}
\end{equation}
where $\ket{+}_{\rm S}$ and $\ket{-}_{\rm S}$ are a pair of orthonormal states obtained by normalizing $\ket{\sqrt\kappa}\pm \ket{-\sqrt\kappa}$.
Since $\lambda'$ is centered around $\eta\mu$ and $R$ is chosen such that 
$\eta\mu R=\eta \kappa$, we have $\alpha_\nu^2\sim \eta \kappa$ whereas
$\tilde{\alpha}^2\sim \kappa$ for $\kappa\ll 1$.
Hence in the B92 protocol, the optical loss decreases the distinguishability of the two signal states sent out from Alice. Eve can exploit this difference to obtain partial information on the bit value, but this time she has no further freedom to manipulate Alice and Bob, since the qubit must be almost always received by Bob. We may thus expect that the key gain is still positive after the privacy amplification even for lossy quantum channels, which will be confirmed in the subsequent sections. 

Another complication in the B92 protocol is that it looks as if we run slightly different S-B92 protocols with parallel qubit channels, for each of which Bob has a filter with a nonorthogonality specified by $\nu$. Since this variation might be exploited by Eve, it is desirable to choose $\lambda'$ to be narrow. As we have seen, on the other hand, $\lambda'$ must be much wider than photon number fractuation ($\sim\sqrt{\eta\mu}$) in the SRP. Due to this trade-off, it is expected that the key rate also depends on the intensity $\mu$ of SRP.

\subsection{The phase error estimation}

Since the formal derivation of the upper bound of the phase error rate is a bit complicated, in the main text we intuitively present the derivation and we leave the formal derivation to Appendices. From the relation to S-B92 we have seen above, it follows that 
Eq.~(\ref{sfinal}) holds for each photon number $\nu$, namely,
\begin{eqnarray}
\Lambda_{{\rm fil},\nu}-2\Lambda_{{\rm bit,\nu}}\le
2G_\nu\alpha_\nu\beta_\nu g(C_\nu{{\bf Z}_\nu})\,.
\end{eqnarray}
 Here, we have used a straightforward translation of the parameters in S-B92 into those in the B92 protocol as 
\begin{eqnarray}
(\Lambda_{{\rm fil}}, \Lambda_{{\rm bit}}, G,\beta,\alpha)\to(\Lambda_{{\rm fil},\nu}, \Lambda_{{\rm bit,\nu}}, G_{\nu},\beta_{\nu},\alpha_{\nu})
\end{eqnarray}
and
\begin{eqnarray}
{\bf Z}&=&(\Lambda_{\rm s}, \Lambda_{1x}, \Lambda_{\rm fil},
\Lambda_{{\rm ph}})^T\nonumber\\
&\to&{\bf Z}_\nu=(\Lambda_{{\rm s},\nu}, \Lambda_{1x,\nu}, \Lambda_{{\rm fil},\nu},
\Lambda_{{\rm ph},\nu})^T
\end{eqnarray}

Note that $g({\bf u})$ is concave and $g$ is monotone
increasing for each element of ${\bf u}$.
Taking the 
summation over the range $\nu\in \lambda$, where
\begin{eqnarray}
\lambda\equiv\lambda^{({\rm D}_1)}\cup\lambda'=[\nu_i,\nu_f]\,,
\label{lambda}
\end{eqnarray}
and noting that 
$G_\nu\alpha_\nu\beta_\nu\le G_{\nu_f}\alpha_{\nu_f}\beta_{\nu_f}$,
we have
\begin{equation}
 \Lambda_{{\rm fil},\lambda}-2\Lambda_{{\rm bit,\lambda}}\le
2G_{\nu_f}\alpha_{\nu_f}\beta_{\nu_f}g(C'{\bf Z}_\lambda)
\label{semifinal}
\end{equation}
with ${\bf Z}_\lambda=(\Lambda_{{\rm s},\lambda}, 
\Lambda_{1x,\lambda}, \Lambda_{{\rm fil},\lambda},
\Lambda_{{\rm ph},\lambda})^T$ and 
$C'\equiv\max_{\nu \in \lambda}C_\nu$, 
where the maximum is taken for each element.

Since we have used Eq.~(\ref{sfinal}) from S-B92 argument \cite{TL04},  
Eq.~(\ref{semifinal}) holds true in the asymptotic limit in which 
the block size for {\em each} photon number $\nu$ is large. As we show in Appendix B, with a more detailed analysis using Azuma's inequality \cite{A67,BTBLR04}, we can show that it holds true if the {\em total} block size is large.

%As we have already mentioned, if we can estimate the upper bound of
%phase error rate, then we finish the security proof. Basically, we can
%estimate the bound thanks to the filter, which is also the case for
%S-B92. The difference is that since we have different filters for each
%parallel qubit channels, we have to consider how to treat the different
%filters. Moreover, recall that we have assumed D$_1$ that can tell if
%the photon number $\nu$ is $\nu\in\lambda^{({\rm
%D}_1)}=[\nu_{i},\nu_{f}-1]$. This means that we lose a perfect
%resolution of each qubit space, and we have to treat them in a unified
%manner. Intuitively, since the function $g$ in Eq.~(\ref{sfinal}) is
%concave and monotone with respect to its arguments, one can obtain the
%inequality for the estimation by taking summation of Eq.~(\ref{sfinal})
%over different qubit space, i.e., the photon number $\nu$. With Azuma's
%inequality \cite{A67,BTBLR04}, we can justify this idea to obtain the
%final expression as follows (see \cite{appendix} for the detail)

Finally, we must convert Eq.~(\ref{semifinal}) into one involving 
only the accessible quantities. As we show in Appendix B, Eqs.~(B10)-(B13), 
${\bf Z}_\lambda$ is upper-bounded
by ${\bf{Z}}_U\equiv (1,  \tilde\alpha^2, 
\Lambda_{\rm fil, {\rm all}},\Lambda_{{\rm ph},\lambda})^T$. 
A lower bound is given 
by ${\bf{Z}}_L\equiv (\tilde\eta_{\lambda}, \tilde\alpha^2-1+\tilde\eta_{\lambda},
\Lambda_{\rm fil,\lambda'},\Lambda_{{\rm ph},\lambda})^T$
with $\tilde\eta_{\lambda}\equiv \Lambda_{{\rm vac},\lambda^{({\rm
D}_1)}}+\Lambda_{{\rm fil,\lambda'}}$. 
Since the vacuum events are dominant in the high loss region, 
$\tilde\eta_{\lambda}$ is almost unity, implying that the qubit loss events
are negligible. This makes our B92 robust against channel losses.
Using these bounds, we can rewrite Eq.~(\ref{semifinal}) into
\begin{equation}
 \Lambda_{{\rm fil},\lambda'}-2\Lambda_{{\rm bit,all}}\le
2G_{\nu_f}\alpha_{\nu_f}\beta_{\nu_f}g(C_{+}'{\bf{Z}}_U-C_{-}'{\bf{Z}}_L)\,,
\label{finalex}
\end{equation}
where we have decomposed $C'=C_{+}'-C_{-}'$ such that 
$C_{+}'$ includes only the nonnegative entries of $C'$. An upper bound of the phase error $\overline{\Lambda_{{\rm ph},\lambda'}}$ is determined by inserting the experimental data into Eq.~(\ref{finalex}) and searching for the maximum value of $\Lambda_{{\rm ph},\lambda}$ satisfying Eq.~(\ref{finalex}), since $\Lambda_{{\rm ph},\lambda'}\le \Lambda_{{\rm ph},\lambda}$ by definition. The key generation rate is given by \cite{SP00} 
\begin{eqnarray}
G_{\rm key}={\rm Max}\Big(0,\Lambda_{{\rm fil},\lambda'}\left[1-h\left(\frac{\Lambda_{{\rm bit},\lambda'}}{\Lambda_{{\rm fil},\lambda'}}\right)-h\left(\frac{\overline{\Lambda_{{\rm ph},\lambda'}}}{\Lambda_{{\rm fil},\lambda'}}\right)\right]\Big)\,, \nonumber\\
\end{eqnarray}
where $h(x)=-x\log_2 x-(1-x)\log_2 (1-x)$.

\section{Key generation rate}
To illustrate the resulting key rates, %we assume that Alice and Bob's devices are noiseless and 
we consider a quantum channel that maps $\hat
P(\ket{\pm\sqrt{\kappa}}_{\rm S}\ket{\sqrt{\mu}}_{\rm SRP})$ into
$(1-p)\hat P(\ket{\pm\sqrt{\eta\kappa}}_{\rm S}\ket{\sqrt{\eta\mu}}_{\rm
SRP})+p\hat P(\ket{1}_{\rm S}\ket{\sqrt{\eta\mu}}_{\rm SRP})$. Here
$\ket{1}_{\rm S}$ is a single photon state and $0\le p\le1$. The first
part models loss with transmission rate $\eta$, and the second part models dark counts since a single photon state causes a random click on Bob's detectors $D_2$ and $D_3$. In order to express the experimental parameters, we define $l$ as the distance between Alice and Bob, $\xi$ as a loss coefficient, and $\eta_{\rm Bob}$ as Bob's detection efficiency. We take the experimental parameters from Gobby et al \cite{GYS04}, which are $p=1.7\times 10^{-6}$, $\xi=0.21$ (db/km), and $\eta_{\rm Bob}=0.045$ neglecting alignment errors. With these parameters, we have
\begin{eqnarray}
\Lambda_{{\rm fil},{\rm all}}&=&e^{-2\eta\kappa}2\eta\kappa(1-p)+e^{-\eta\kappa}p\,, \nonumber\\
\eta&=&10^{-\frac{\xi l}{10}}\eta_{\rm Bob}\,, \nonumber\\
\Lambda_{\rm fil,\lambda'}&=&\Lambda_{{\rm fil},{\rm all}}\sum_{\nu\in\lambda^{\rm (D_1)}}P_{\eta(\mu-\kappa)}(\nu)\,,\nonumber\\
\Lambda_{{\rm vac},\lambda^{({\rm D}_1)}}&=&e^{-2\eta\kappa}(1-p)\sum_{\nu\in\lambda^{\rm(D_1)}}P_{\eta(\mu-\kappa)}(\nu)\,,\nonumber\\
\Lambda_{{\rm bit},\lambda'}&=&\frac{e^{-\eta\kappa}p}{2}\sum_{\nu\in\lambda^{\rm
(D_1)}}P_{\eta(\mu-\kappa)}(\nu)\,,
\end{eqnarray}
where $P_x(\nu)\equiv e^{-x}\frac{x^{\nu}}{\nu!}$.
%%%%%%%%%%%%%%%%%%%%%%%%%%%%%%%%%%%%%%%%%%%%%%%%figure3
\begin{figure}[tbp]
\begin{center}
 \includegraphics[scale=0.6]{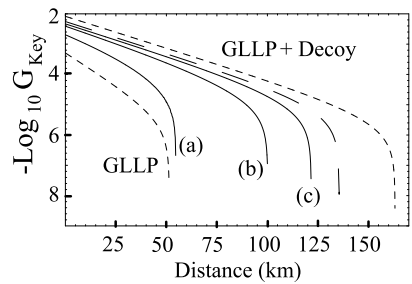}
\end{center}
 \caption{The key generation rate. See the main text for the explanations.
\label{key}}
\end{figure}
%%%%%%%%%%%%%%%%%%%%%%%%%%%%%%%%%%%%%%%%%%%%%%%%

We have calculated the key rate $G_{\rm key}$ and the achievable distance $l_{\rm a}$ (km) for several choices 
of parameters $(\mu, \kappa, a)$, 
where $\nu_i=\eta\mu-a\sqrt{\eta\mu}$, and $\nu_f=\eta\mu+a\sqrt{\eta\mu}$.
In Fig.~\ref{key}, we fix $\kappa=10^{-0.92}$ and $a=3.2$, and vary
$\mu$ as (a) $\mu=10^5$ resulting in $l_{\rm a}=55$,
(b) $\mu=10^{6.59}$ $(l_{\rm a}=100)$, and 
(c) $\mu=10^{10}$ $(l_{\rm a}=122)$.
 This confirms our earlier speculation that 
increasing $\mu$ will restrict Eve's options and
leads to a better key rate.
The slope of curve (c) shows that the key rate is proportional 
to the channel transmission $\eta$ until the distance approaches 
$l_{\rm a}$.
%In (b), we have chosen the minimum value of $\mu$ to achieve 100 km.
Increasing $\mu$ further beyond (c) merely results in saturation
at $l_{\rm a}=124$, and this is the maximum
distance among the combinations of the parameter set that we have tried. We can also trade the intensity of SRP for a 
poor resolution $a$. For example, with $\mu=10^7$ we can still achieve $l_a=100$ even with a relative resolution 
$|\lambda^{({\rm D}_1)}|/\eta\mu\sim 10\%$.

To see how accurate our phase estimation is, we plot 
$G_{\rm key}$ (dashed
line) based on the actual phase error rate induced by the quantum
channel. The curve is almost identical for (a), (b) and (c), 
implying that the key rate dependence on $\mu$ is due to 
the difference in the estimation ability.
This is a typical feature of the B92 protocol in which the phase error rate cannot be estimated directly
as opposed to the BB84 protocol. We must also note that 
the derivation of Eq.~(\ref{finalex}) from Eq.~(\ref{sfinal})
is not tight and the estimation might be improved by
a more sophisticated analysis.

For comparison, we have also shown the key rate for the BB84 protocol
based on the GLLP formula \cite{GLLP02}
(dotted line, $l_{\rm a}=51$),
and the rate with infinite number of decoy states \cite{W03}
(dotted line, $l_{\rm a}=163$), assuming the ideal error correction 
efficiency. 
Considering that the analysis of the statistical fluctuations in the
decoy schemes tends to be very complicated \cite{W03}, 
achieving a comparably long distance with just two states 
could be an advantage of the B92 protocol in practical implementations.

\section{Summary}
In summary, we studied the security of the original B92 with
strong reference pulse assuming the two types of detectors. We can identify qubit spaces composed of the states with the signal pulse including zero and one photon. It follows that even if the transmission
 channel is physically very lossy, Bob still finds a qubit state with high probability, 
which is the essential difference from the single-photon B92. 
We showed the key rate scales as $G_{\rm key}\sim O(\eta)$ when 
the SRP is strong enough. It is interesting to remove the assumptions on the detectors, which we leave for future studies.

\section{acknowledgement}

Part of this work was performed when K.T worked for QIT Group, Universit$\ddot{\mbox{a}}$t Erlangen-N$\ddot{\mbox{u}}$rnberg in Germany, the Perimeter Institute, and the University of Toronto in Canada. This work was in part supported by JSPS Grant-in-Aid for Scientific Research(C) 20540389. We thank D. Gottesman, H-K. Lo, and G. Weihs for valuable discussions and supports. N.L acknowledges the support by the German Research Council (DFG) through an Emmy-Noether Group, the European Project SECOQC and NSERC (Discovery Grant, QuantumWorks).

\appendix
\section{Azuma's inequality}
In the following appendices, we explain the formal derivation of the phase error estimation, Eq. (\ref{finalex}). As we have mentioned in the main text, we need to obtain the relationship among $\Lambda_{{\rm bit}, \lambda}$, $\Lambda_{{\rm fil}, \lambda}$, $\Lambda_{{\rm ph}, \lambda}$, $\Lambda_{1_x, \lambda}$, and $\Lambda_{{\rm s}, \lambda}$. For the derivation of the relationship, we have two problems to be addressed. First, note that these ratios cannot be directly measured in the experiment. For instance, in the actual experiment we have direct access to $\Lambda_{{\rm fil}, \lambda'}$, not $\Lambda_{{\rm fil}, \lambda}$. Thus, $\Lambda_{{\rm fil}, \lambda}$ has to be estimated via actually available experimental data. This is also the case for other ratios, such as $\Lambda_{{\rm bit}, \lambda}$, $\Lambda_{1_x, \lambda}$, and $\Lambda_{{\rm s}, \lambda}$, and the estimation will be discussed in Appendix C. 

In addition to this problem, we have another difficulty in deriving the relationship. The difficulty lies in the fact that not all the observables can be simultaneously measured. For example, $\Lambda_{1_x, \lambda}$ and $\Lambda_{{\rm bit}, \lambda}$ do not commute because Alice has to conduct X-basis measurement for $\Lambda_{1_x, \lambda}$ while she performs Z-basis measurement for $\Lambda_{{\rm bit}, \lambda}$. In order to overcome this problem, we randomly label each pair as a test pair
(with a small probability $t$) or a code pair (with probability $1-t$) \cite{quant2}. We try to distill a key from the code pairs, and hence what we need
to estimate is the phase error rate $\Lambda_{{\rm ph}, \lambda}$ in the code pairs. By using the Bell measurement \cite{LC98}, $\Lambda_{{\rm bit}, \lambda}$ and $\Lambda_{{\rm ph}, \lambda}$ in the code pairs can be measured simultaneously, in principle. On the other hand, for the test pairs,
we assume that Alice performs X-basis measurement on her qubits
to measure $\Lambda_{1_x, \lambda}$. In this way, all the observables
can be measured simultaneously, and we can clearly define
joint probabilities. This is where Azuma's inequality, which
connects conditional probabilities with the
actual ratios, comes into our game.

In order to claim that the above fictitious senario is possible
in principle, the actual protocol must be slightly modified.
The labeling of test pairs and code pairs should be publicly done
in the actual protocol. The key is distilled only from the code
pairs, while Alice discards her bit
values for the test pairs, such that she could have performed
X-basis measurement on her fictitious qubits.
The only parameters we actually need to monitor for the test pairs
is the number of events where zero or one photon are recorded by
$D_2$ and $D_3$ in total, with the outcome of $D_1$ being in the valid range
$\lambda_{D_1}$. This quantity is used in Eq. (C5) below.

\begin{figure}[tbp]
\begin{center}
 \includegraphics[scale=0.45]{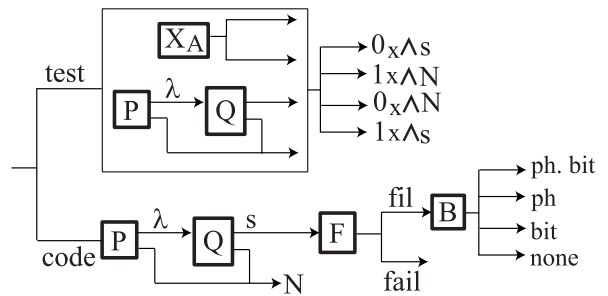}
\end{center}
 \caption{The measurement diagram.
\label{diagram}}
\end{figure}

Now let us describe our argument for the phase error estimation
precisely. Imagine a sequence of measurements on a pair of systems, A
and (S,SRP), which Alice and Bob share (see also Fig.~\ref{diagram}). At the beginning, Alice randomly decides whether a pair is a test pairs (with
probability $t$) or a code pairs (with probability $1-t$), and
for the code pair, the sequence of measurements starts 
with ``P'' with outcome $\nu$, followed by
projection ``Q''. If P fails to detect photon number inside $\lambda$ or
Q does not succeed in projecting the state to the qubit state ${\cal H}^{(\nu)}$, we call such outcome as ``N''. If the state 
was found in the qubit state with P detecting photon number inside $\lambda$ (we call this outcome ``{\rm s}''),
we subject the state to the filtering operation F. If F succeeds (``fil''),
we apply the Bell measurement \cite{nielsen} (``B'') to see whether there is 
a bit error (``bit'') and/or a phase error (``ph'').
If F fails (``fail''), we simply discard it.

On the other hand, for the test pairs Alice performs $X$-basis measurement on ${\cal H}_A$, and Bob conducts the aforementioned ``P'' measurement and the 
projection ``Q''. As a result, we have four possible outcomes, $0_x \wedge{\rm s}$,
$1_x \wedge {\rm s}$, $0_x \wedge {\rm N}$, and $1_x \wedge {\rm N}$.

%we randomly subject the state (with probability $0<t<1$) to Alice's $X$-basis measurement on ${\cal H}_A$ (``$1_x$'') to obtain ``A0'' or ``A1'' or subject to the filtering operation F (with probability $1-t$). We call the pairs that are subjected to the $X$-basis measurement as ``test'' pairs and the other as ``code'' pairs. If F succeeds (``fil''), we apply the Bell measurement \cite{nielsen} (``B'') to see whether there is a bit error (``bit'') and/or a phase error (``ph''). If F fails (``fail''), we simply discard it. Each set of outcomes from these measurements corresponds to a ``path'' in the diagram of Fig.~\ref{key}.

%%%%%%%%%%%%%%%%%%%%%%%%%%%%%%%%%%%%%%%%%%%%%%%%
%%%%%%%%%%%%%%%%%%%%%%%%%%%%%%%%%%%%%%%%%%%%%%%%

Consider $N$ such pairs of systems, and imagine that
 we repeat this set of measurements from the $1^{st}$ pair in order.
Let $\xi^{(k)}$ be the path that was actually taken by the $k^{th}$ pair.
For $\Omega\in\{{\rm bit}, {\rm ph}, {\rm fil}, 1_x\wedge{\rm s}, {\rm {\rm s}}\}$, define $\Upsilon_{\Omega, \nu}^{(l)}$ as the number of the pairs whose path include $\{\Omega,\nu\}$ among the first $l$ pairs. Furthermore, we define $\Upsilon_{\Omega, \lambda}^{(l)}$ as $\sum_{\nu\in\lambda}\Upsilon_{\Omega, \nu}^{(l)}$, and specifically we write
$\Upsilon_{\Omega, \lambda}\equiv\Upsilon_{\Omega, \lambda}^{(N)}$. With these parameters, we define
\begin{eqnarray}
\Lambda_{\Omega,\lambda}&\equiv&\frac{\Upsilon_{\Omega,\lambda}}{N(1-t)}\,\,\,{\rm for}\,\,\Omega={\rm s}, {\rm fil}, {\rm ph}, {\rm bit}\nonumber\\
\Lambda_{\Omega,\lambda}&\equiv&\frac{\Upsilon_{\Omega,\lambda}}{Nt}\,\,\,{\rm for}\,\,\Omega=1_x\wedge{\rm s}\,.
\end{eqnarray}

Our task is to derive a bound on $\Lambda_{{\rm ph}, \lambda}$
as a function of other $\Lambda_{\Omega, \lambda}$'s that can 
be estimated from the data available in the actual protocol.
To do so, we invoke the fact that we can assign a joint probability 
 $P_{\rm path}(\{\xi^{(k)}\}_{k=1,\ldots N})$ for every attack 
by Eve, since all $\{\xi^{(k)}\}$ can be measured at the same time.
Then, we can apply a known classical theorem to $P_{\rm path}$.
From $P_{\rm path}$, define conditional 
probability $\varsigma_{\Omega, \nu}^{(l)}$ for $\xi^{(l)}$ to include $\{\Omega,\nu\}$ conditioned on
$\{\xi^{(k)}\}_{k=1,\ldots,l-1}$, and especially we define $\varsigma_{\Omega, \lambda}^{(l)}\equiv\sum_{\nu\in\lambda}\varsigma_{\Omega, \nu}^{(l)}$ and $\varsigma_{\Omega, \lambda}\equiv\varsigma_{\Omega, \lambda}^{(N)}$. Furthermore, we introduce $n_{\Omega,\nu}$ as follows
\begin{eqnarray}
n_{\Omega,\lambda}&\equiv&\frac{\varsigma_{\Omega,\lambda}}{1-t}\,\,\, {\rm for}\,\,\Omega={\rm s}, {\rm fil}, {\rm bit}, {\rm ph}\nonumber\\
n_{\Omega,\lambda}&\equiv&\frac{\varsigma_{\Omega,\lambda}}{t}\,\,\,{\rm for}\,\,\Omega=1_x\wedge{\rm s}\,,
\end{eqnarray}
where $n_{\Omega,\lambda}\equiv\sum_{\nu\in\lambda}n_{\Omega,\nu}$. Then, Azuma's inequality \cite{A67} states that, since 
$X_{\Omega, \lambda}^{(l)}\equiv \Upsilon_{\Omega,
\lambda}^{(l)}-\sum_{k=1}^{l}\varsigma_{\Omega, \lambda}^{(k)}$ 
is martingale \cite{martingale} and satisfies the bounded difference condition \cite{BDC},
it follows that for any $\epsilon>0$ and $N>0$, 
\begin{eqnarray}
{\rm Prob}\left(|\Upsilon_{\Omega,
\lambda}-\sum_{l=1}^{N} \varsigma_{\Omega,
\lambda}^{(l)}|>N\epsilon\right)\le2 e^{-\frac{N\epsilon^2}{2}}
\label{Azumafirst}
\end{eqnarray}
holds \cite{BTBLR04} for $\Omega={\rm s}, {\rm fil}, {\rm ph}, {\rm bit}, 1_x\wedge{\rm s}$. 

%By the same token, for $\Omega=1_x\wedge{\rm s}$
%\begin{eqnarray}
%{\rm Prob}\left(|\Upsilon_{\Omega,
%\lambda}-\sum_{l=1}^{N} \varsigma_{\Omega,
%\lambda}^{(l)}|>N\epsilon\right)\le2 e^{-\frac{N\epsilon^2}{2}}\,
%\label{Azumasecond}
%\end{eqnarray}
%holds for any $\epsilon>0$ and $N>0$.

Then, for the code pairs, by setting $\epsilon$ in Eq.~(\ref{Azumafirst}) as $\epsilon(1-t)$, we have
\begin{eqnarray}
{\rm Prob}\left(|\Lambda_{\Omega,\lambda}-\sum_{l=1}^{N} n_{\Omega,
\lambda}^{(l)}/N|>\epsilon\right)\le2 e^{-\frac{N(1-t)^2\epsilon^2}{2}}\,,
\label{finite1}
\end{eqnarray}
where $\Omega={\rm s}, {\rm fil}, {\rm ph}, {\rm bit}$, and similarly for the test pairs we have
\begin{eqnarray}
{\rm Prob}\left(|\Lambda_{\Omega,\lambda}-\sum_{l=1}^{N} n_{\Omega,
\lambda}^{(l)}/N|>\epsilon\right)\le2 e^{-\frac{Nt^2\epsilon^2}{2}}\,,
\label{azumathird}
\label{finite2}
\end{eqnarray}
where $\Omega=1_x\wedge{\rm s}$. These inequalities guarantee that in the limit of large $N$ with $t$ being fixed, the normalized version of the sum of the probabilities $n_{\Omega, \lambda}^{(l)}/N$ differs from the actual ratio only with exponentially small probability. The next step is to find the relationship among the probabilities $n_{\Omega, \lambda}^{(l)}$ so that we have the relationship among the number of the actual events.

Let $\hat\rho^{(l)}$ be the density operator of the $l^{th}$ pair conditioned on the previous outcomes $\{\xi^{(k)}\}_{k=1,\ldots,l-1}$. Then $n_{\Omega, \lambda}^{(l)}$ can be expressed as $n_{\Omega, \lambda}^{(l)}={\rm Tr}\left(\hat\rho^{(l)}\hat{F}_{\Omega,\lambda}\right)$, where POVM element $\hat{F}_{\Omega,\lambda}\equiv\sum_{\nu\in\lambda}\hat{F}_{\Omega,\nu}$ is defined as follows.
\begin{eqnarray}
\hat{F}_{{\rm s},\nu}&\equiv&{\hat{\bf 1}}_{\rm A}\otimes\hat{\bf 1}_{\rm B}^{(\nu)}\nonumber\\
\hat{F}_{1_x\wedge{\rm s}, \nu}&\equiv&P(\ket{1_x}_{\rm A})\otimes\hat{\bf 1}_{\rm B}^{(\nu)}\nonumber\\
\hat{F}_{{\rm fil},\nu}&\equiv&\hat{\bf 1}_{\rm A}\otimes{\hat A}_{\rm s}^{(\nu)\dagger}{\hat A}_{\rm s}^{(\nu)}\nonumber\\
\hat{F}_{{\rm bit},\nu}&\equiv&\hat P(\ket{0_z}_{\rm A})\otimes\hat F_{1}^{(\nu)}+\hat P(\ket{1_z}_{\rm A})\otimes\hat F_{0}^{(\nu)}\nonumber\\
\hat{F}_{{\rm ph},\nu}&\equiv&\hat P(\ket{0_x}_{\rm A})\otimes\hat P({\hat A}_{\rm s}^{(\nu)\dagger}\ket{1_x^{(\nu)}}_{\rm B})\nonumber\\
&+&\hat P(\ket{1_x}_{\rm A})\otimes\hat P({\hat A}_{\rm s}^{(\nu)\dagger}\ket{0_x^{(\nu)}}_{\rm B})\,.
\end{eqnarray}
Here, $\hat{\bf 1}_{\rm A}\equiv\hat P(\ket{0_z}_{\rm A})+\hat P(\ket{1_z}_{\rm A})$, $\hat{\bf 1}_{\rm B}^{(\nu)}\equiv\hat P(\ket{0_z^{(\nu)}}_{\rm B})+\hat P(\ket{1_z^{(\nu)}}_{\rm B})$, and see Eqs.~(\ref{Bit value}) and (\ref{filter}) for the definitions of ${\hat F}_i^{(\nu)}$ and ${\hat A}_{\rm s}^{(\nu)}$, respectively.

Having $\hat\rho^{(l)}$ and $\hat{F}_{\Omega,\lambda}$, now we can write down
$n_{\Omega, \lambda}^{(l)}={\rm Tr}(\hat\rho^{(l)}\hat{F}_{\Omega,\lambda})$
by using the corresponding POVM element $\hat{F}_{\Omega,\lambda}$. We have little clue
about the identity of $\hat\rho^{(l)}$, but we may find 
a relation $f(\{n_{\Omega, \lambda}^{(l)}\}_{\Omega})\le0$ 
%with $n_{\Omega, \lambda}\equiv{\rm Tr}(\hat\rho\hat{F}_{\Omega,\lambda})/\zeta_{\Omega}$,
which holds for any state $\hat\rho^{(l)}$. If we find such 
a convex function $f$, it follows that 
$f(\{\sum_{l=1}^{N}n_{\Omega, \lambda}^{(l)}/N\}_{\Omega})\le 0$,
and we obtain a relation among $\{\Lambda_{\Omega, \lambda}\}$
through Azuma's inequality.
Note that in the limit of $N\rightarrow\infty$, the above relation is simply 
$f(\{\Lambda_{\Omega, \lambda}\}_{\Omega})\le0$. 

Thus, the derivation of the relation among the conditional probabilities is the
key point in deriving the inequality for the phase error estimation, which we
present in Appendix B.

%We note that the ``test'' in the entanglement distillation protocol corresponds to Alice and Bob who just discard some portion of signals in the B92 protocol. Since arbitrary small number of ``test'' is enough for having the phase error estimation, the fraction of the discard is negligible in the B92 protocol.

\section{Detailed calculations for the phase error estimation}
In this appendix, we derive the relation among the conditional probabilities $n_{\Omega, \lambda}^{(l)}={\rm Tr}\left(\hat\rho^{(l)}\hat{F}_{\Omega,\lambda}\right)$. In what follows, we assume $\nu<1/R$ ($R$ is the reflectivity of the beam splitter BS1) and $\nu_f\le\frac{-1}{2\ln (1-R)}$. We note that in the actual experiment, these assumptions are well-justified, and even if these assumptions do not hold, we can construct the relationship by applying slight modifications. Thus, these assumptions are not essential for the proof.

Since the relationship we derive in this appendix holds for any density matrix, we use the abbreviation $n_{\Omega, \lambda}={\rm Tr}\left(\hat\rho\hat{F}_{\Omega,\lambda}\right)$, where $\rho$ is any density operator. Let us introduce $m_{ii',\nu}$ and $n_{ii',\nu}$ $(i,i'=0,1)$ as ${\rm Tr}\left(\hat\rho\hat{P}(\ket{\Gamma_{ii'}^{(\nu)}}_{\rm AB})\right)$ and ${\rm Tr}\left(\hat\rho\hat{P}(\ket{i_x}_{\rm{A}}\ket{{i'}_{x}^{(\nu)}}_{\rm B})\right)$, respectively. Here, $\ket{\Gamma_{ii'}^{(\nu)}}_{\rm AB}\equiv(-1)^{ii'}\beta_{\nu}\ket{i_x}_{\rm{A}}\ket{{i'}_{x}^{(\nu)}}_{\rm B}+(-1)^{i(i'+1)}\alpha_{\nu}\ket{(i+1)_x}_{\rm{A}}\ket{({i'+1})_{x}^{(\nu)}}_{\rm B})$ (the summation and multiplication are taken in modulo 2).

If we regard the subspace spanned by $\{\ket{0_x}_{\rm{A}}\ket{{0}_{x}^{(\nu)}}_{\rm B}, \ket{1_x}_{\rm{A}}\ket{{1}_{x}^{(\nu)}}_{\rm B}\}$ ($\{\ket{0_x}_{\rm{A}}\ket{{1}_{x}^{(\nu)}}_{\rm B}, \ket{1_x}_{\rm{A}}\ket{{0}_{x}^{(\nu)}}_{\rm B}\}$) as a qubit, then the two bases, $\{\ket{0_x}_{\rm{A}}\ket{{0}_{x}^{(\nu)}}_{\rm B}, \ket{1_x}_{\rm{A}}\ket{{1}_{x}^{(\nu)}}_{\rm B}\}$ and $\{\ket{\Gamma_{00}^{(\nu)}}_{\rm AB}, \ket{\Gamma_{11}^{(\nu)}}_{\rm AB}\}$, correspond to two directions in the Bloch sphere with relative angle $\theta^{(\nu)}$ satisfying 
$\alpha_{\nu}^{2}=\sin^2\theta^{(\nu)}$, and $\beta_{\nu}^{2}=\cos^2\theta^{(\nu)}$  in $[0,\pi/2]$. Then, for any qubit state in the
Bloch sphere, we have
\begin{eqnarray}
\sin^2(\theta_{0}^{(\nu)}-\theta^{(\nu)})&\le& \sin^2\phi_{0}^{(\nu)} \le
\sin^2(\theta_{0}^{(\nu)}+\theta^{(\nu)})\,, \nonumber\\
\label{bloch}
\end{eqnarray}
where $n_{11,\nu}/(n_{11,\nu}+n_{00,\nu})=\sin^2\theta_{0}^{(\nu)}$ and $m_{11,\nu}/(m_{11,\nu}+m_{00,\nu})=\sin^2\phi_{0}^{(\nu)}$.
Similarly we have
\begin{eqnarray}
\sin^2(\theta_{1}^{(\nu)}-\theta^{(\nu)})&\le& \sin^2\phi_{1}^{(\nu)} \le
\sin^2(\theta_{1}^{(\nu)}+\theta^{(\nu)})\,, \nonumber\\
\label{bloch1}
\end{eqnarray}
where $n_{01,\nu}/(n_{01,\nu}+n_{10,\nu})=\sin^2\theta_{1}^{(\nu)}$ and $m_{01,\nu}/(m_{01,\nu}+m_{10,\nu})=\sin^2\phi_{1}^{(\nu)}$.

In addition, $n_{\Omega, \nu}$ can be expressed as 
\begin{eqnarray}
n_{{\rm {\rm s}},\nu}&=&\sum_{i,i'=0,1}n_{ii',\nu}\label{ns}\\
n_{1_x\wedge{\rm s}, \nu}&=&(n_{10,\nu}+n_{11,\nu})
\label{n1x}\\
n_{{\rm fil},\nu}&=&G_{\nu}[\alpha_{\nu}^{2}(n_{00,\nu}+n_{10,\nu}\label{nfil}\nonumber\\
&+&\beta_{\nu}^{2}(n_{01,\nu}+n_{11,\nu})]\\
n_{{\rm ph},\nu}&=&G_{\nu}\left(\alpha_{\nu}^{2}n_{10,\nu}+\beta_{\nu}^{2}n_{01,\nu}\right)\label{nph}\\
n_{{\rm bit}, \nu}&=&\frac{G_{\nu}}{2}(m_{11,\nu}+m_{01,\nu})\label{nbit}\,.
\end{eqnarray}

Note that Eqs. (\ref{ns})-(\ref{nph}) can be represented as
\begin{equation}
{\bf Z}_{\nu}\equiv\left(
\begin{array}{cccc}
n_{{\rm {\rm s}},\nu} \\
n_{1_x\wedge{\rm s}, \nu} \\
n_{{\rm fil},\nu}\\
n_{{\rm ph},\nu}
\end{array}
\right)=C_{\nu}^{-1}\left(
\begin{array}{cccc}
n_{00,\nu} \\
n_{01,\nu} \\
n_{10,\nu}\\
n_{11,\nu}
\end{array}
\right)\,,
\label{matrixeq}
\end{equation}
where
\begin{equation}
C_{\nu}^{-1}=\left(
\begin{array}{cccc}
1&1 &1 &1 \\
0&0 &1 &1 \\
G_{\nu}\alpha_{\nu}^{2}&G_{\nu}\beta_{\nu}^{2} &G_{\nu}\alpha_{\nu}^{2} &G_{\nu}\beta_{\nu}^{2} \\
0&G_{\nu}\beta_{\nu}^{2} &G_{\nu}\alpha_{\nu}^{2} & 0
\end{array}
\right)\,.
\end{equation}

By solving Eqs. (\ref{nbit}), (\ref{bloch}), and (\ref{matrixeq}), we have an inequality
\begin{eqnarray}
& &n_{{\rm fil},\nu}-2n_{{\rm bit},\nu}\le2\,G_{\nu}\alpha_{\nu}\beta_{\nu}g(C_\nu{{\bf Z}_\nu})\,.\nonumber\\
\label{eq1}
\end{eqnarray}
Here, $g((a,b,c,d)^T)\equiv \sqrt{ad}+\sqrt{bc}$ and $C_\nu$ is given by
\begin{equation}
C_{\nu}=\left(
\begin{array}{cccc}
\frac{\beta_{\nu}^4}{(\beta_{\nu}^2-\alpha_{\nu}^2)}&-\beta_{\nu}^2 &-\frac{\alpha_{\nu}^2}{G_{\nu}(\beta_{\nu}^2-\alpha_{\nu}^2)} &-\frac{1}{G_{\nu}} \\
-\frac{\alpha_{\nu}^4}{(\beta_{\nu}^2-\alpha_{\nu}^2)}&-\alpha_{\nu}^2 &\frac{\alpha_{\nu}^2}{G_{\nu}(\beta_{\nu}^2-\alpha_{\nu}^2)} &\frac{1}{G_{\nu}} \\
\frac{\alpha_{\nu}^2\beta_{\nu}^2}{(\beta_{\nu}^2-\alpha_{\nu}^2)}&\beta_{\nu}^2 &-\frac{\beta_{\nu}^2}{G_{\nu}(\beta_{\nu}^2-\alpha_{\nu}^2)} &\frac{1}{G_{\nu}} \\
-\frac{\alpha_{\nu}^2\beta_{\nu}^2}{(\beta_{\nu}^2-\alpha_{\nu}^2)}&\alpha_{\nu}^2 &\frac{\beta_{\nu}^2}{G_{\nu}(\beta_{\nu}^2-\alpha_{\nu}^2)} &-\frac{1}{G_{\nu}}
\end{array}
\right)\,.
\end{equation}

We can overestimate $C_\nu{{\bf Z}_\nu}$ by maximizing each entry of $C_{\nu}$ over $\nu\in\lambda=[\nu_i, \nu_f]$ to obtain $C'\equiv {\rm max}_{\nu\in\lambda}C_{\nu}$ as
\begin{equation}
C'=\left(
\begin{array}{cccc}
\frac{\beta_{\nu f}^4}{(\beta_{\nu f}^2-\alpha_{\nu f}^2)}&-\beta_{\nu f}^2 &-\frac{\alpha_{\nu i}^2}{G_{\nu i}(\beta_{\nu i}^2-\alpha_{\nu i}^2)} &-\frac{1}{G_{\nu i}} \\
-\frac{\alpha_{\nu i}^4}{(\beta_{\nu i}^2-\alpha_{\nu i}^2)}&-\alpha_{\nu i}^2 &\frac{\alpha_{\nu f}^2}{G_{\nu f}(\beta_{\nu f}^2-\alpha_{\nu f}^2)} &\frac{1}{G_{\nu f}} \\
\frac{\alpha_{\nu f}^2\beta_{\nu f}^2}{(\beta_{\nu f}^2-\alpha_{\nu f}^2)}&\beta_{\nu i}^2 &-\frac{\beta_{\nu i}^2}{G_{\nu i}(\beta_{\nu i}^2-\alpha_{\nu i}^2)} &\frac{1}{G_{\nu f}} \\
-\frac{\alpha_{\nu i}^2\beta_{\nu i}^2}{(\beta_{\nu i}^2-\alpha_{\nu i}^2)}&\alpha_{\nu f}^2 &\frac{\beta_{\nu f}^2}{G_{\nu f}(\beta_{\nu f}^2-\alpha_{\nu f}^2)} &-\frac{1}{G_{\nu i}}
\end{array}
\right)\,.
\label{Cmatrix}
\end{equation}
Here, we have used the assumption that we choose $\lambda$ such that $\beta_{\nu}^2>\alpha_{\nu}^2$, i.e., $\nu R<1$, for any $\nu\in\lambda$, leading to $\partial_{\nu}\beta_{\nu}^2\le0$, $\partial_{\nu}\alpha_{\nu}^2\ge0$, $\partial_{\nu}G_{\nu}\le0$, $\partial_{\nu}\frac{\alpha_{\nu}^4}{(\beta_{\nu}^2-\alpha_{\nu}^2)}\ge0$, $\partial_{\nu}\frac{\beta_{\nu}^4}{(\beta_{\nu}^2-\alpha_{\nu}^2)}\ge0$, $\partial_{\nu}\frac{\alpha_{\nu}^2\beta_{\nu}^2}{(\beta_{\nu}^2-\alpha_{\nu}^2)}\ge0$, $\partial_{\nu}\frac{\alpha_{\nu}^2}{(\beta_{\nu}^2-\alpha_{\nu}^2)}\ge0$, and $\partial_{\nu}\frac{\beta_{\nu}^2}{(\beta_{\nu}^2-\alpha_{\nu}^2)}\ge0$. Furthermore, the condition $\nu_f\le\frac{-1}{2\ln (1-R)}$ gives us $\partial_{\nu}G_{\nu}\alpha_{\nu}\beta_{\nu}\ge0$, and with the help of this inequality, we can slightly modify Eq.~(B9) as
\begin{eqnarray}
n_{{\rm fil},\nu}-2n_{{\rm bit}, \nu}\le2G_{\nu_f}\alpha_{\nu_f}\beta_{\nu_f}g(C'{{\bf Z}_\nu})\,.
\label{eq2}
\end{eqnarray}
Note that if $\nu R<1$ and $\nu_f\le\frac{-1}{2\ln (1-R)}$ do not hold, it just changes the sign of the derivative functions. Thus, even in that case, we can construct a matrix similar to Eq.~(\ref{Cmatrix}) by applying appropriate modifications, such as the interchange of $\nu_{i}$ and $\nu_{f}$. Another remark is on our assumption $\nu_f<\frac{1}{R}$ and
$\nu_f\le -\frac{1}{2\ln(1-R)}$. Note that when $R\ll1$, this condition is approximately equivalent to $\frac{1}{2R}=\frac{\mu}{2\kappa}\sim10\mu\gg\nu_f$, which is natural for standard experiments where Bob almost never detects the photon number that is greater than the mean photon number emitted by Alice due to the channel loss and non-unit quantum efficiency of the detectors. Thus, in most of normal experiments, the assumptions hold.

Finally, we take summation over $\nu\in\lambda$, and with the help of the concavity of the function $g((a,b,c,d)^T)$, we have
\begin{eqnarray}
& &n_{{\rm fil},\lambda}-2n_{{\rm bit},\lambda}\nonumber\\
&\le& 2G_{\nu_f}\alpha_{\nu_f}\beta_{\nu_f}\sum_{\nu\in\lambda}g(C'{{\bf z}_\nu})\nonumber\\
&\le& 2G_{\nu_f}\alpha_{\nu_f}\beta_{\nu_f}g(C'\sum_{\nu\in\lambda}{{\bf z}_\nu})\nonumber\\
&=&2G_{\nu_f}\alpha_{\nu_f}\beta_{\nu_f}g(C'{{\bf z}_\lambda})\,,
\label{eqeq2}
\end{eqnarray}
where
\begin{equation}
{\bf z}_{\lambda}%\equiv\left(\begin{array}{cccc}\sum_{\nu\in\lambda}n_{{\rm {\rm s}},\nu} \\\sum_{\nu\in\lambda}n_{1_x\wedge{\rm s}, \nu} \\\sum_{\nu\in\lambda}n_{{\rm fil},\nu}\\\sum_{\nu\in\lambda}n_{{\rm ph},\nu}\end{array}\right)
=\left(
\begin{array}{cccc}
n_{{\rm {\rm s}},\lambda} \\
n_{1_x\wedge{\rm s},\lambda} \\
n_{{\rm fil},\lambda}\\
n_{{\rm ph},\lambda}
\end{array}
\right)\,.
\end{equation}

\section{Estimation of the parameters from the experimentally available data}

In this appendix, we first express Eq.~(\ref{eqeq2}) in terms of the actual number of ratio, and then we give the estimation of the parameters appearing in the resulting inequality to obtain Eq.~(\ref{finalex}). First, we apply Azuma's inequality to ${\bf z}_{\lambda}$. In the limit of large $N$, ${\bf z}_{\lambda}$ is transformed into the actual ratio of the corresponding events ${\bf Z}_{\lambda}^T\equiv (\Lambda_{{\rm {\rm s}},\lambda}, \Lambda_{1_x\wedge{\rm s},\lambda}, \Lambda_{{\rm fil},\lambda}, \Lambda_{{\rm ph},\lambda})^T$, and we have
\begin{eqnarray}
\Lambda_{{\rm fil},\lambda}-2\Lambda_{{\rm bit},\lambda}\le2G_{\nu_f}\alpha_{\nu_f}\beta_{\nu_f}g(C'{{\bf Z}_\lambda})\,,
\label{eq2}
\end{eqnarray}
whereas for finite $N$, we can bound the probability of violating (a slightly relaxed version of) Eq.~(\ref{eq2}) using Eq.~(\ref{finite1}) and Eq.~(\ref{finite2}). 

The next step is to give the estimation of the variables appearing in 
Eq.~(\ref{eq2}) from the actually observed quantities in the experiment. By considering an inclusion relation on $\lambda$, %For the explanation, we introduce $\Theta^{\lambda^{({\rm D}_1)}}(c)$ as the actual ratio of event that ${\rm D}_1$ detects photons inside $\lambda^{({\rm D}_1)}$ and $c$ photons are detected by ${\rm D}_2$ and ${\rm D}_3$ in total. With this notation, 
$\Lambda_{{\rm fil},\lambda}$ can obviously be bounded as
\begin{eqnarray}
\Lambda_{{\rm fil},\lambda'}\le\Lambda_{{\rm fil},\lambda}\le\Lambda_{{\rm fil}, {\rm all}}\nonumber\\
\end{eqnarray}
where we have used $\lambda'\in\lambda$ (see Eq.~(\ref{lambda})). Similarly, $\Lambda_{{\rm bit},\lambda}$ is bounded as
\begin{eqnarray}
\Lambda_{{\rm bit},\lambda'}\le\Lambda_{{\rm bit},\lambda}&\le&\Lambda_{{\rm bit},{\rm all}}\,,
\end{eqnarray}
where we have defined $\Lambda_{{\rm bit},{\rm all}}$ as the number of the bit errors divided by $N(1-t)$ and $\Lambda_{{\rm bit},\lambda'}$ as that of the bit errors divided by $N(1-t)$ with the condition $\nu\in\lambda'$.

Next, note that $\Lambda_{{\rm s},\lambda}$ is the ratio of events of the successful qubit projection with the total photon number inside $\lambda$. This ratio is lower-bounded by the ratio of event $\Lambda_{{\rm vac},\lambda^{({\rm D}_1)}}$ where ${\rm D}_1$ detects photons inside $\lambda^{({\rm D}_1)}$ and the vacuum is detected by ${\rm D}_2$ and ${\rm D}_3$ in total, plus the ratio of event $\Lambda_{{\rm fil,\lambda'}}$ where ${\rm D}_1$ detects photons inside $\lambda^{({\rm D}_1)}$ and a single-photon is detected by ${\rm D}_2$ and ${\rm D}_3$ in total. Thus, we have
\begin{eqnarray}
& &\tilde\eta_{\lambda}\equiv\Lambda_{{\rm vac},\lambda^{({\rm D}_1)}}+\Lambda_{{\rm fil,\lambda'}}\le\Lambda_{{\rm s},\lambda}\le1\,.\nonumber\\
\label{tildeeta}
\end{eqnarray}

Finally, $\Lambda_{1_x\wedge{\rm s},\lambda}$ is obviously upper-bounded by $\Lambda_{1_x}$ that is the ratio for Alice to obtain $1_x$ in the test pairs. On the other hand, $\Lambda_{1_x\wedge{\rm s},\lambda}$ is lower-bounded from the worst case scenario where Alice's X-basis measurement results in 1 for all the pairs that have resulted in Bob's ``N''. The ratio of this event is represented by $1-\Lambda_{{\rm s},\lambda}^{(t)}$, where $\Lambda_{{\rm s},\lambda}^{(t)}$ is the test-pair-version of $\Lambda_{{\rm s},\lambda}$. By the same token as Eq.~(\ref{tildeeta}), $1-\Lambda_{{\rm s},\lambda}^{(t)}$ is no larger than $1-\tilde\eta_{\lambda}^{(\rm t)}$ where
\begin{eqnarray}
\tilde\eta_{\lambda}^{(\rm t)}\equiv\Lambda_{{\rm vac},\lambda^{({\rm D}_1)}}^{(\rm t)}+\Lambda_{{\rm fil,\lambda'}}^{(\rm t)}\,
\end{eqnarray}
is the test-pair-version of $\tilde\eta_{\lambda}$, and $\Lambda_{{\rm vac},\lambda^{({\rm D}_1)}}^{(\rm t)}$ and $\Lambda_{{\rm fil,\lambda'}}^{(\rm t)}$ are the test-pair-version of $\Lambda_{{\rm vac},\lambda^{({\rm D}_1)}}$ and $\Lambda_{{\rm fil,\lambda'}}$, respectively. Hence, we have
\begin{eqnarray}
\Lambda_{1_x}-(1-\tilde\eta_{\lambda}^{(\rm t)})\le\Lambda_{1_x\wedge{\rm s},\lambda}\le\Lambda_{1_x}\,.
\label{1x}
\end{eqnarray}
Moreover, by applying the substitutions $\Lambda_{\Omega, \lambda}\rightarrow\Lambda_{1_x}$ and $n_{\Omega,\lambda}^{(l)}\rightarrow\tilde\alpha^2$ to Eq.~(\ref{azumathird}), we have 
\begin{eqnarray}
{\rm Prob}\left(|\Lambda_{1_x}-\tilde\alpha^{2}|>\epsilon\right)\le 2e^{-N(t\epsilon)^2/2}
\end{eqnarray}
for any $N>0$ and $\epsilon>0$ (see Eq.~(\ref{alpha}) for the definition of $\tilde\alpha$). Thus, we can bound $\Lambda_{1_x\wedge{\rm s},\lambda}$ by using the experimentally available data as
\begin{eqnarray}
\tilde\alpha^2-(1-\tilde\eta_{\lambda}^{(\rm t)})\le\Lambda_{1_x\wedge{\rm s},\lambda}\le\tilde\alpha^2\,,
\label{1xfinal}
\end{eqnarray}
which is violated with probability less than $2e^{-N(t\epsilon)^2/2}$.

% according to Bernoulli trial, $\Lambda_{1_x}$ exponentially approaches to the probability Alice has $1_x$, i.e., $\frac{1-e^{-2\kappa}}{2}\equiv \tilde\alpha^2$. More precisely, for large $N$ and $\epsilon>0$ \begin{eqnarray} &&{\rm Prob}\left(|\Lambda_{1_x}tN-\tilde\alpha^{2}tN|>\epsilon tN\right)\le e^{-2N(t\epsilon)^2}\nonumber\\ &\Leftrightarrow&{\rm Prob}\left(|\Lambda_{1_x}-\tilde\alpha^{2}|>\epsilon\right)\le e^{-2NM(t\epsilon)^2} \end{eqnarray} holds, leading that \begin{eqnarray} \tilde\alpha^2-(1-\tilde\eta_{\lambda})\le\Lambda_{1_{x}\wedge{\rm s},\lambda}\le \tilde\alpha^2 \end{eqnarray} is also violated with probability asymptotically less than $e^{-2N(t\epsilon)^2}$ \cite{Bernoulli}.

In summary, we have ${\bf{Z}}_L\le{\bf Z}_{\lambda}\le{\bf{Z}}_U$, where ${\bf{Z}}_L\equiv (\tilde\eta_{\lambda}, \tilde\alpha^2-1+\tilde\eta_{\lambda}^{(t)},
\Lambda_{\rm fil,\lambda'},\Lambda_{{\rm ph},\lambda})^T$ and ${\bf{Z}}_U\equiv (1,  \tilde\alpha^2, \Lambda_{\rm fil, {\rm all}},\Lambda_{{\rm ph},\lambda})^T$, and the final expression is described as
\begin{equation}
 \Lambda_{{\rm fil},\lambda'}-2\Lambda_{{\rm bit,all}}\le
2G_{\nu_f}\alpha_{\nu_f}\beta_{\nu_f}g(C_{+}'{\bf{Z}}_U-C_{-}'{\bf{Z}}_L)\,,
\label{finalappendix}
\end{equation}
where we have decomposed $C'=C_{+}'-C_{-}'$ such that 
$C_{+}'$ includes only the nonnegative entries of $C'$. We remark that Eq.~(\ref{finalappendix}) uses two parameters $\tilde\eta_{\lambda}^{(t)}$ and $\tilde\eta_{\lambda}$, whereas Eq.~(\ref{finalex}) uses only $\tilde\eta_{\lambda}$. Since $\tilde\eta_{\lambda}^{(t)}$ and $\tilde\eta_{\lambda}$ are the same parameter in different random samples,
they must be very close to each other for large $N$. Hence it is also possible
to measure only $\tilde\eta_{\lambda}$ in the code pairs and estimate $\tilde\eta_{\lambda}^{(t)}$ from that. But in practice, there is no additional overhead in determining $\tilde\eta_{\lambda}^{(t)}$, and thus
it is always better to use Eq.~(\ref{finalappendix}) involving fewer numbers of estimations.


\begin{thebibliography}{99}
\bibitem{martingale} A sequence of random variables $X^{(0)}, X^{(1)}, ..., X^{(s+1)}$ is called a martingale iff the expectation of $X^{(s+1)}$ conditional on a realisation of the sequence of random variables $X^{(0)}, X^{(1)}, ..., X^{(s)}$ is equal to $X^{(s)}$, i.e., $E[X^{(s+1)}|X^{(s)},X^{(s-1)},\cdots, X^{(0)}]=X^{(s)}$. Note that $E[X^{(s+1)}|X^{(s)},X^{(s-1)},\cdots, X^{(0)}]$ can be regarded as a function of $X^{(0)}, X^{(1)}, ..., X^{(s)}$. One can show that $X_{\Omega, \lambda}^{(s)}$ is martingale as follows. First note that 
\begin{eqnarray}
X_{\Omega, \lambda}^{(s+1)}=\xi_{\Omega, \lambda}^{(s+1)}-\varsigma_{\Omega, \lambda}^{(s+1)}+X_{\Omega, \lambda}^{(s)}\,,
\end{eqnarray}
where $\xi_{\Omega, \lambda}^{(s+1)}$ is a random raviable satisfying $\xi_{\Omega, \lambda}^{(s+1)}=1$ iff we have the event $\{\Omega, \lambda\}$, and $\xi_{\Omega, \lambda}^{(s+1)}=0$ otherwise. Then, we have
\begin{eqnarray}
& &E[X_{\Omega, \lambda}^{(s+1)}|X_{\Omega, \lambda}^{(s)}, X_{\Omega, \lambda}^{(s-1)},\cdots, X_{\Omega, \lambda}^{(0)}]\nonumber\\
&=&E[\xi_{\Omega, \lambda}^{(s+1)}|X_{\Omega, \lambda}^{(s)}, X_{\Omega, \lambda}^{(s-1)},\cdots, X_{\Omega, \lambda}^{(0)}]\nonumber\\
&-&E[\varsigma_{\Omega, \lambda}^{(s+1)}|X_{\Omega, \lambda}^{(s)}, X_{\Omega, \lambda}^{(s-1)},\cdots, X_{\Omega, \lambda}^{(0)}]\nonumber\\
&+&E[X_{\Omega, \lambda}^{(s)}|X_{\Omega, \lambda}^{(s)}, X_{\Omega, \lambda}^{(s-1)},\cdots, X_{\Omega, \lambda}^{(0)}]\nonumber\\
&=&\varsigma_{\Omega, \lambda}^{(s+1)}-\varsigma_{\Omega, \lambda}^{(s+1)}+X_{\Omega, \lambda}^{(s)}\nonumber\\
&=& X_{\Omega, \lambda}^{(s)}\nonumber\,.
\end{eqnarray}
\bibitem{BDC} A sequence of random variables $X^{(0)}, X^{(1)}, ..., X^{(s)}$ is said to satisfy the bounded difference condition (BDC) if $|X^{(s)}-X^{(s-1)}|\le c_s$ holds for any $s$ and positive $c_s$. In our case, it is obvious that $|X_{\Omega, \lambda}^{(s)}-X_{\Omega, \lambda}^{(s-1)}|\le1$.
\end{thebibliography}
\end{document}